\documentclass[twoside,a4paper]{article}
\usepackage{amsmath,graphicx,amssymb,fancyhdr,amsthm,enumerate,textcomp} 
\usepackage[usenames]{color}
\newtheorem{thm}{Theorem}[section]

\theoremstyle{definition} 
\newtheorem{defn}[thm]{Definition}
  
\theoremstyle{remark}  
  
\def\beq{\begin{eqnarray}}  
\def\eeq{\end{eqnarray}}  
\def\bsp{\begin{split}}  
\def\esp{\end{split}}

\def\d{\mathrm{d}}

\def\VC{V^{\mathbb{C}}}
\def\GC{G^{\mathbb{C}}}

\newcommand{\mf}[1]{{\mathfrak #1}}   
   
\newcommand{\mb}[1]{{\mathbb #1}}   
   
\newcommand{\mbold}[1]{\mbox{\boldmath{\ensuremath{#1}}}}

\begin{document}   
   
\title{\Large\textbf{All metrics have curvature tensors characterised by its invariants as a limit: the $\epsilon$-property.}}  
\author{{\large\textbf{Sigbj\o rn Hervik  }    }
 \vspace{0.3cm} \\    
Faculty of Science and Technology,\\    
 University of Stavanger,\\  N-4036 Stavanger, Norway    \\
\vspace{0.3cm} \\     
\texttt{sigbjorn.hervik@uis.no} }    
\date{\today}    
\maketitle  
\pagestyle{fancy}  
\fancyhead{} 
\fancyhead[EC]{S. Hervik}  
\fancyhead[EL,OR]{\thepage}  
\fancyhead[OC]{The epsilon property}  
\fancyfoot{} 
  
\begin{abstract}
We prove a generalisation of the $\epsilon$-property, namely that for any dimension and signature, a metric which is not characterised by its polynomial scalar curvature invariants, there is a frame such that the components of the curvature tensors can be arbitrary close to a certain ``background''. This ``background''  is defined by its curvature tensors: it is characterised by its curvature tensors and has the same polynomial curvature invariants as the original metric.
\end{abstract}
\section{Introduction}
For Lorentzian spacetimes having all vanishing polynomial curvature invariants (VSI spaces) it has been proven that the $\epsilon$-property holds \cite{epsilon}. The $\epsilon$-property implies the components of the curvature tensors can, by chosing a suitable frame, be arbitrarily small. Clearly, this property can only hold for VSI spaces as the invariants are all zero. In the general case however, in particular for degenerate metrics, we will here prove a similar property but at the cost of having to replace the $\epsilon$-property with respect to a ``background''. This is remincent of the fact that any matrix $M$ can be split, using the Jordan decomposition, into a diagonalisable matrix, $D$, and a nilpotent matrix, $N$: 
\[ M=D+N.\] 
The components of the nilpotent matrix can, by a change of basis, be made as small as possible. Also note that the matrices $M$ and $D$ have the same polynomial invariants: $\mathrm{tr}(M^n)=\mathrm{tr}(D^n)$. Therefore, since the eigenvalues (=diagonal components) of $D$, are determined by the polynomial invariants $\mathrm{tr}(D^n)$ we can say that $D$ is ``characterised by its invariants''. For a tensor this concept can be defined in a similar way \cite{procesi, eberlein, operator,align}. 

In order for us to state the corresponding result for curvature tensors, we will review some results from invariant theory and  define the appropriate concepts which we need. Furthermore,  we will consider the polynomials invariants and so in what follows 'invariants' is to be understood as '\emph{polynomial invariants}'. 

The idea is to consider a group $G$ acting on a vector space $V$. In our case we will be consider a real $G$ and a real vector space $V$. However, it is advantageous to review the complex case with a complex group $\GC$ acting on a complex vector space $\VC$.  Then for a vector $X\in \VC$ we can define the \emph{orbit} of $X$ under the action of $\GC$ as follows: 
\[ \mathcal{O}_{\mathbb{C}}(X)\equiv \{  g(X)\in\VC ~\big{|}~g\in \GC \}\subset \VC\]

Then (\cite{procesi}, p555-6): 
\begin{thm}
If $\GC$ is a linearly reductive group acting on an affine variety $\VC$, then the ring of invariants is finitely generated. Moveover, the quotient $\VC/ \GC$ parameterises the closed orbits of the $\GC$-action on $\VC$ and the invariants separate closed orbits.   
\end{thm}
Here the term \emph{closed} refers to \emph{topologically closed} with respect to the standard vector space topology and henceforth, closed will mean topologically closed. 
This implies that given two distinct closed orbits $A_1$ and $A_2$, then there is an invariant with value $1$ on $A_1$ and $0$ on $A_2$. This enables us to define the set of orbits: 
\beq
\mf{C}_{\mathbb{C}}=\{\mathcal{O}_{\mathbb{C}}(X)\subset \VC ~\big{|}~ \mathcal{O}_{\mathbb{C}}(X) \text{ closed.}  \}
\eeq
Based on the above theorem we can thus say that the invariants separate elements of $\mf{C}_{\mb{C}}$ and hence we will say that an element of  $\mf{C}_{\mb{C}}$ is \emph{characterised by its invariants}. 

In our case we will consider the real case where we have the Lorentz group, $O(1,n-1)$ which is a real semisimple group. For real semisimple groups acting on a real vector space we do not have the same uniqueness result as for the complex case \cite{eberlein}. However, by complexification, $[G]^{\mathbb{C}}=\GC$ we have $[O(1,n-1)]^{\mathbb{C}}=O(n,\mathbb{C})$, and complexification of the real vectorspace $V$ we get $\VC\cong V+iV$. The complexification thus lends itself to the above theorem, and consequently we will define \emph{characterised by its invariants} as follows. For a tensor, $T$, a rotation of a frame naturally defines a group action on the \emph{components} of the tensor. Then:
\begin{defn}
Consider a (real) tensor, $T\in V$, or a direct sum of tensors, then if the orbit of the components of $T$ under the complexified Lorentz group $G^{\mathbb{C}}$ is an element of $\mf{C}_{\mb{C}}$, i.e., $\mathcal{O}_{\mathbb{C}}(T) \in \mf{C}_{\mathbb{C}}$, then we will say that $T$ is \emph{characterised by its invariants}. 
\end{defn}
\noindent As the invariants parameterise the set $\mf{C}_{\mb{C}}$ and since the group action defines an equivalence relation between elements in the same orbit this definition makes sense.

Let us similarly define the real orbits: 
\[\mathcal{O}(X)\equiv \{ g(X)\in V ~\big{|}~g\in G \}\subset V\]
How do these results translate to the real case? The real orbit $\mathcal{O}(T)$, is a real section of the complex orbit  $\mathcal{O}_{\mathbb{C}}(T)$. However, there might be more than one such real section having the same complex orbit. Using the results of \cite{eberlein}, these real closed orbits are disjoint, moreover: 
\begin{thm}
$\mathcal{O}(T)$ is closed in $V$ $\Leftrightarrow$ $\mathcal{O}_{\mathbb{C}}(T)$ is closed in $\VC$.
\end{thm}
Thus the question of whether $T$ is characterised by its invariants is thus equivalent to whether $\mathcal{O}(T)$ is closed in $V$. Thus we can define similarly: 
\beq
\mf{C}=\{\mathcal{O}(X)\subset V ~\big{|}~ \mathcal{O}(X) \text{ closed.}  \},
\eeq
hence, we have that $T$ is characterised by its invariants iff $\mathcal{O}(T)\in\mf{C}$. 

However, as pointed out, there might be other closed real orbits  $\mathcal{O}(\tilde{T})$ having the same invariants as  $\mathcal{O}({T})$ (in line with the comments in \cite{inv,operator}). An example of this is the pair of metrics:
\beq
\d s^2_1&=& -\d t^2+\frac{1}{x^2}\left(\d x^2+\d y^2+\d z^2\right),\nonumber \\
\d s^2_2&=& \d \tau^2+\frac{1}{x^2}\left(\d x^2+\d y^2-\d \zeta^2\right),
\eeq
The curvature tensors\footnote{Both of these metrics are symmetric and conformally flat, so the only non-zero curvature tensor is the Ricci tensor.}  of these metrics lie in separate orbits $\mathcal{O}(T)$, but in the same complex orbit $\mathcal{O}_{\mathbb{C}}(T)$.  

\section{The $\epsilon$-property}
We introduce a basis  of orthonomal (or null) vectors $\omega=\{ {\mbold e}_1,...,{\mbold e}_n\}$ for an $n$-dimensional pseudo-Riemannian space. We express the components of the curvature tensors up to $k$th derivatives, with respect to the frame $\omega$, in terms of the vector $X_{\omega}=[R_{abcd},R_{abcd;e},...,R_{abcd;e_1...e_k}]$. At a point $p$ on the manifold this vector can be considered as $X_{\omega}\in \mathbb{R}^m$, for some $m$. We will also use the standard Euclidean norm $||X||$ in $\mathbb{R}^m$. 

We note that there is no requirement that the components of $R_{abcd}$ etc. need to be independent (nor is the order significant). One way this can be thought of is that we have a metric from which we compute the Riemann tensors with respect to an orthonormal frame $\omega$. We when construct the $m$-tuple $X_{\omega}$ from these components. These components may be dependent (which would be the case if we just naively write down all components of Riemann without considering the symmetries). If the metric is of signature $(q,n-q)$, then the action of the corresponding orthogonal group $O(q,n-q)$ will preserve these symmetries and thus this has no consequence for our result. 

The aim of this Note is to prove that (assuming $C^\infty$ and that the algebraic structure does not change over a neighbourhood\footnote{This is a technical assumption implying that the Segre/Petrov or Ricci/Weyl types cannot change over the neighbourhood. In some cases this assumption can be relaxed, for example for real analytic metrics, or if we are only interested in a \emph{pointwise} $\epsilon$-property, however we will not consider this here.}): 
\begin{thm}[$\epsilon$-property]
Consider a spacetime $(\mathcal{M},g)$ of any dimension (and signature) and a fixed number of derivatives, $k$, of the Riemann tensor. Then locally either:
\begin{enumerate}
\item{} The curvature tensors $X_\omega$  are characterised by its invariants; or,
\item{} For any $\epsilon>0$, there exists a suitable frame and an $\tilde{X}_{\omega}\in \mathbb{R}^m$ such that the components $X_{\omega}=\tilde{X}_{\omega}+N_{\omega}$, where $||N_{\omega}||<\epsilon$ and $\tilde{X}_{\omega}$ is characterised by its invariants. Moreover, the polynomial invariants of $\tilde{X}_{\omega}$ are the same as those of $X_{\omega}$.
\end{enumerate}
\end{thm}

We will prove this theorem using the orbits of $X_\omega$ under frame rotations. However, in the 4 dimensional Lorentzian spacetime there is a much simpler proof following along the same lines as in the VSI case. 
\begin{proof}[Proof: 4D Lorentzian case] 
Here we can use the results of \cite{inv,Kundt} which implies that either the spacetime is $\mathcal{I}$-non-degenerate, hence characterised by its invariants, or degenerate Kundt. For a degenerate Kundt the vector $X_{\omega}$ can be written as $X_\omega=(X_\omega)_0+(\text{negative boost weight terms})$. Thus we can set $\tilde{X}_{\omega}=(X_\omega)_0$ (which is characterised by its invariants) and $N_{\omega}$ is the negative boost weight part. By applying a boost we can thus get the components of  $N_{\omega}$ as small as possible, in particular $||N_{\omega}||<\epsilon$. Clearly also the polynomial invariants of $\tilde{X}_{\omega}$ are the same as those of $X_\omega$. 
\end{proof}
It is believed that a similar mechanism can be used in abitrary dimensions and signatures (for Lorentzian case, see \cite{align}). However, in order to give a proof in the general case we will use a different method using the space of orbits of $X_\omega$. In order to do this we will review some known results from invariant theory. 

A frame-rotation in $n$-dimensions (arbitrary signature) at a point $P\in\mathcal{M}$ is an action of the group $g\in O(q,n-q)$. A rotation of the frame $g\omega=\{ M^a_{~1}{\mbold e}_a,..., M^a_{~n}{\mbold e}_a\}$, induces an action of $g$ on $X_{\omega}$ through the tensor structure of the components:
\[ g(X_{\omega})=\left[ M^{b_1}_{~a_1}...M^{b_{k_1}}_{~a_{k_1}}R^{(1)}_{b_1...b_{k_1}},M^{b_1}_{~a_1}...M^{b_{k_2}}_{~a_{k_2}}R^{(2)}_{b_1...b_{k_2}},...,M^{b_1}_{~a_1}...M^{b_{k_N}}_{~a_{k_N}}R^{(N)}_{b_1...b_{k_N}}\right]. \]
 Consider now the orbit of a point $X\in\mathbb{R}^m$: 
\[ \mathcal{O}(X)=\{ g(X)|g\in  O(q,n-q)\}\subset \mathbb{R}^m. \] 
Clearly, if $X$ and $Y$ are in the same orbit, then they would be equivalent as curvature tensors because they are separated by a mere rotation of frame. 
The equivalence problem is then reduced to a question of classifying the various orbits. 

In the complex case, where we consider the complexification of $O(q,n-q)\mapsto [O(q,n-q)]^{\mathbb{C}}=O(n,\mathbb{C})$ and $\mathbb{R}^m\mapsto \mathbb{C}^m$, the orbits that are characterised by their invariants are the closed orbits (those that are elements of $\mf{C}_{\mathbb{C}}$). In our case we do not consider orbits under the the complex group $O(n,\mathbb{C})$, rather one of the real forms  $O(q,n-q)$ for which we need to consider, as discussed in the introduction, the set of real orbits $\mf{C}$.  

Assume thus that $\mathcal{O}(X)\in \mf{C}$; i.e., that $\mathcal{O}(X)$ is closed and from the previous discussion that $ \mathcal{O}(X)$ is characterised by its invariants. Indeed this implies that there exists a $\tilde{X}\in \mathcal{O}(X)$ which is \emph{minimal} \cite{eberlein}. Recall that a vector $\tilde{X}\in\mb{R}^m$ is \emph{minimal} iff $||g(\tilde{X})||\geq ||\tilde{X}||$ for all $g\in G$ where is $|| \cdot ||$  the $O(q)\times O(n-q)$-invariant Euclidean norm on $\mb{R}^m$.  There may be several minimal vectors of each orbit but the minimal vectors are in some sense the vectors with the ``smallest'' norm with respect to the Euclidean metric on $\mb{R}^m$. The existence of minimal vectors are not essential for this discussion, however, it is useful to note that the limit in the theorem can always be chosen to be a minimal vector \cite{eberlein}. In particular, the existence of a minimal $\tilde{X}$ in the orbit of $X$ is equivalent to saying that $\mathcal{O}(X)$ is closed and hence that $X$ is  characterised by its invariants \cite{eberlein}.

We are now ready to complete the proof:
\begin{proof}[Proof: General case]
Assume that $X_\omega$ has a closed orbit; i.e., $\mathcal{O}(X_\omega)$ is closed. Due to the existence of a minimal $\tilde{X}$ in its orbit it is characterised by its invariants by the above argument.  

Assume thus that  $\mathcal{O}(X_\omega)$ is not closed. Then by the results of \cite{eberlein}, there is a unique closed orbit in the closure of  $\overline{\mathcal{O}(X_\omega)}$. In particular, there would be a minimal $\tilde{X}_{\omega}$ in $\overline{\mathcal{O}(X_\omega)}$. Since  $\tilde{X}_{\omega}$ is in the closure of $\mathcal{O}(X_\omega)$, there would exist, for any $\epsilon>0$ a sequence $x_n\in \mathcal{O}(X_\omega)$ and an integer $N$ so that $||x_n-\tilde{X}_{\omega}||<\epsilon$ for all $n>N$. Thus since the sequence, $x_n$, is in the orbit of $X_\omega$, we can choose a frame such that $x_n=X_\omega$ for an $n>N$; hence, there  exists a frame such that $||X_\omega-\tilde{X}_{\omega}||<\epsilon$. 

As regards to the polynomial invariants, $I_i$, we note that these are continuous as functions $I_i: ~\mathbb{R}^m\mapsto \mathbb{R}$. Since the invariants are constants over the orbit $\mathcal{O}(X_\omega)$, they must also, due to continuity, be constants over the closure $\overline{\mathcal{O}(X_\omega)}$. Consequently, $\tilde{X}_{\omega}$ and $X_\omega$ have the same polynomial invariants. This completes the proof.
\end{proof}

Let us also make some final remarks on this result. 

Firstly, there is no guarantee that the minimal vector $\tilde{X}_\omega$ corresponds to the curvature tensors of an actual metric but there are important cases where it does. For example, in the VSI case $\tilde{X}_\omega=0$, and for sufficiently large $k$, this would be flat space. For other spacetimes, for example, CSI spacetimes there is a frame such that $\tilde{X}_\omega$ is constant over the neighbourhood. 

As an example, consider the Weyl (Petrov) type III vacuum solution of Robinson-Trautman \cite{RT}:
\[ \d s^2=-\d t^2+t^2\d x^2+t^{\frac
25}\left(e^{-\frac{\sqrt{6}}{5}x}\d y+\frac{\sqrt{5}}2t^{\frac 45}\d
x\right)^2+t^{\frac 65}e^{\frac{4\sqrt{6}}{5}x}\d z^2.\]
 This has all vanishing zeroth order invariants (VSI$_0$); hence, including only the Riemann tensor, $X^{(0)}=[R_{abcd}]$, we have  $\tilde{X}^{(0)}=0$. The first order invariants, however, are non-zero. In particular, $\nabla (\text{Riem})$ is of type II, so defining  $X^{(1)}=[R_{abcd},R_{abcd;e}]$, then both $\tilde{X}^{(1)}$ and $N^{(1)}$ are non-zero. However, consider the second derivatives then $X^{(2)}=[R_{abcd},...,R_{abcd;e_1e_2}]$ is characterised by its invariants. Thus this is an example where the $\epsilon$-property depends on the number of derviatives considered. 

Also note that the theorem is not a ``uniform'' $\epsilon$-property in the sense that we do need to truncate $X_{\omega}$ so that it contains only a finite number of derivatives of the Riemann tensor. Thus the validity of an $\epsilon$-property for an infinite number of derivatives is still elusive and remains an open question.\footnote{On the same token,  Cartan \cite{Cartan} showed that, as far as  the equivalence problem is concerned, it is sufficient to know the covariant derivatives $\nabla^{(q)}({\rm Riem})$ up to a certain order $q$ (see also \cite{MP} which considers 4D Lorentzian manifolds). }

We would also point out that the actual limit arises from the existence of a boost: by boosting the frame appropriately, this limit can be achieved (see \cite{align} for the Lorentzian case). Physically, this implies that an observer in, for example, an AdS-gyraton \cite{gyraton} can experience the space arbitrarily close to AdS space. Interestingly, similar spacetimes are of particular interest when it comes to supersymmetry  \cite{susy} and the $\epsilon$-property thus puts these solutions and their physical significance in new light.

This Note also gives a set of new ideas and techniques to study the relationship between the polynomial invariants and the curvature tensors. In particular, these ideas can also be used to study more closely the alignment of degenerate tensors \cite{align}.

\end{document}